\documentclass{article}
\usepackage{spconf,amsmath,graphicx}

\usepackage{adjustbox}
\usepackage{cite}
\usepackage{amsmath,amssymb,amsfonts}
\usepackage{algorithmic}
\usepackage{graphicx}
\usepackage{textcomp}
\usepackage{hyperref}
\usepackage{soul}
\usepackage{color}
\usepackage{booktabs}

\title{Improving Robustness using Joint Attention Network For Detecting Retinal Degeneration From Optical Coherence Tomography Images}
\name{Sharif Amit Kamran $^\star$ \qquad Alireza Tavakkoli $^\star$ \qquad Stewart Lee Zuckerbrod $^\dagger$}
\address{$^\star$ Department of Computer Science and Engineering, University of Nevada, Reno, NV 89557\\
$^\dagger$ Houston Eye Associates, Houston, TX 77401} 
\begin{document}
%
\maketitle

\begin{abstract}
Noisy data and the similarity in the ocular appearances caused by different ophthalmic pathologies pose significant challenges for an automated expert system to accurately detect retinal diseases. In addition, the lack of knowledge transferability and the need for unreasonably large datasets limit clinical application of current machine learning systems. To increase robustness, a better understanding of how the retinal subspace deformations lead to various levels of disease severity needs to be utilized for prioritizing disease-specific model details. In this paper we propose the use of disease-specific feature representation as a novel architecture comprised of two joint networks -- one for supervised encoding of disease model and the other for producing attention maps in an unsupervised manner to retain disease specific spatial information. Our experimental results on publicly available datasets show the proposed joint-network significantly improves the accuracy and robustness of state-of-the-art retinal disease classification networks on unseen datasets.

\end{abstract}


\section{INTRODUCTION}
\label{sec:introduction}
Age-related neuroocular diseases cause irreversible vision loss in about 10\% of the population in the United States and world-wide. It is estimated that blindness and vision loss increase threefold for each decade over 40 in developed countries. In particular, Age-related macular degeneration (AMD) affects over 8.7\% of the population worldwide, and is among the leading causes of blindness \cite{wong2014global}. Although, there have been notable advancements in the anti-angiogenesis therapy, providing patients with treatment options that can slow the progression of the disease \cite{lim2012age}, without early detection neuroocular diseases result in permanent vision loss. 

Optical Coherence Tomography (OCT) is a technique used for imaging the retinal subspace and its different layer structures \cite{sri2014}. Using this technique, sub-retinal tissue formation is encoded and retrieved as data by using  back-scattered  light  by  spectral analysis \cite{mlSD-OCT2017}. OCT images are among tools used by ophthalmologists to diagnose a variety of age-related eye diseases. Due to human error and the lack of clinical consensus for certain eye disease criteria, mis-categorization is common. The major factor contributing to mis-categorization is the stark resemblance between AMD and other retinal degenerative neuroocular diseases 
\cite{yau2012global}. For example, in chorodial neovascularization (CNV), the advance stage of AMD, new blood vessels sprouts up by breaking from the Bruch membrane into the sub–retinal pigment epithelium (sub-RPE) or sub-retinal space. Experts often find it troublesome to distinguish between the severity of AMD and CNV. 

Machine learning and image processing techniques have been actively employed to develop expert systems for neurodegenerative ocular diseases to some degree of success. However, robustness and transferability of these procedures are still questionable. To address this problem, we propose a joint classification and segmentation architecture. 

\section{Related Work}
To identify and predict retinal degenerative diseases simple yet rigorous image processing techniques were used by extracting features, contrast stretching, and thresholding \cite{baghaie2015state}. One such approach for segmenting multiple retinal boundaries fuses recognition and delineation to detect degeneration, anomalies and diseases from cross sectional Retinal OCT images \cite{debuc2011review}. Similar techniques using global optimization methods such graph cuts and region based delineation are also developed to detect different anomalies and degeneration throughout the retina~\cite{vermeer2011automated} and for diagnosing thickness of the chorodial maps and neovascularization~\cite{alonso2013automatic,philip2016choroidal}. These early techniques achieved recall scores of around 80\% for detecting Diabeitc Macular Edema (DME)~\cite{sanchez2004retinal, ege2000screening}. It has been suggested that expanding and enlarging layer density of retina is effective in identifying Diabetic Retinopathy (DR)~ \cite{mlSD-OCT2017}. Segmentation based approaches have been utilized for diagnosing underlying causes of liquid formation in the intra-retinal subspace by detecting abnormal retinal features and comparing the differences between healthy and the degenerated retinal tissue~\cite{meindertniemeijer2012,quellec2010three,lee2010segmentation}. Although exploiting segmentation of retinal layers as an initial step to carry out identification of retinal degeneration has shown promise, this step results in severe inaccuracies when applied on OCT images in the wild - i.e., images acquired from different types of OCT systems~\cite{kafieh2013review}.

Recently, architectures using convolutional neural networks (CNNs) have received the attention of the ophthalmology community, due to their achieving high degrees of precision in identifying different retinal degenerative diseases~\cite{lee2017deep}. Fang et al. combined deep learning with graph search to automatically segment retinal layer boundaries and degeneration for patients having AMD\cite{fang2017automatic}, while Xu et al. developed a Dual-stage framework that exploits deep learning to segment pigment epithelium detachment \cite{xu2017dual}. Despite the use of large collection of images to learn, almost all deep learning architectures are notorious in lacking robustness to detect retinal degeneration in the wild. Therefore, an effective system capable of robustly retaining spatial information of different diseases for improving the detection accuracy in the wild is needed. 

In this paper, we propose a novel joint-attention-network mechanism that can be attached to any classification architecture to improve its robustness and effectiveness for identifying retinal diseases in the wild. In order to accomplish this, we propose a novel dual adaptive loss that can be fine-tuned as a hyper-parameter to prioritize either of the loss values to further improve classification results from the spatial information identified by the attention network. Furthermore, the proposed work exploits both supervised and unsupervised learning by using the proposed dual loss for weight and gradient updates. Our experimental results provide evidence that the proposed joint-attention-network applied to a number of traditional architectures significantly improve the precision, recall, and accuracy of the architecture in the wild. This is clinically significant as ophthalmologists can add the proposed network to any CNN-based architecture trained on any dataset and utilize the overall network on their OCT images collected in their clinical practice without the need to retrain the network to avoid significant performance drops.

\begin{figure}[t]
    \centering
    \includegraphics[width=.7\columnwidth]{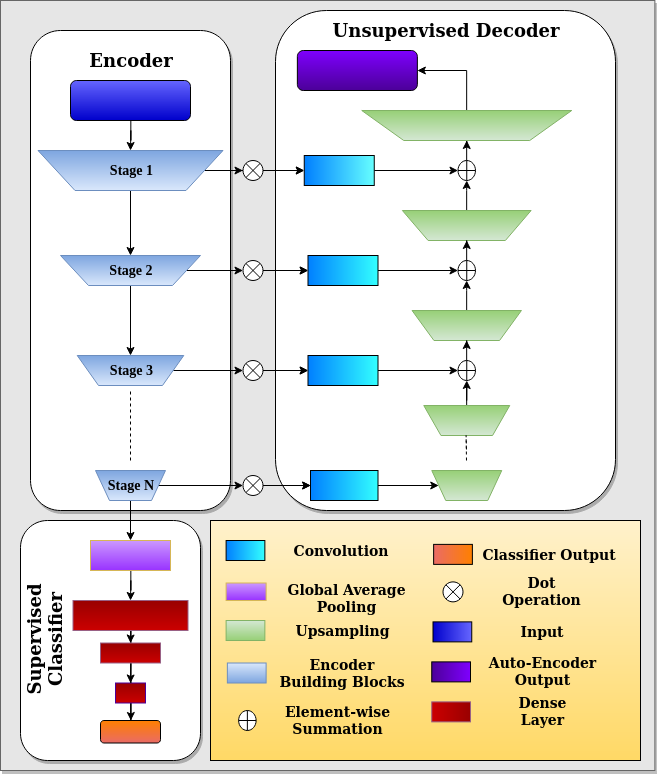}
    \caption{Joint network consisting of (1) Supervised Encoder (2) Unsupervised Decoder and (3) Supervised Classifier.}\vspace{-.15in}
    \label{backbone_architecture}
\end{figure}

\section{Methodology}
The aim of recent CNN architectures for image classification, such as ResNet-50 \cite{he2016deep}, Xception \cite{chollet2017xception}, and MobileNet \cite{sandler2018mobilenetv2}, is to increase accuracy while keeping the number of parameters low. Recent improvements on lowering the network parameters of these architectures have been proposed, including, Pre-activation \cite{he2016identity}, depth-wise convolution \cite{kaiser2017depthwise}, and interspersed atrous separable convolution and regular convolution \cite{kamran2019optic}. However, the robustness of these architectures to performing in the wild has not been fundamentally tackled. This is evident by the poor performance of these networks when presented by unseen data on which the network is not validated. The high dependency of supervised algorithms on training data leads to lower accuracy of the model in the wild. To address this problem we combine both supervised and unsupervised learning into a join network. The following section explains the architecture in detail.

\subsection{Joint-network for Supervised and Unsupervised Learning}
Fig.~\ref{backbone_architecture} shows the overall architecture of the proposed join-attention-network. Given any classification architecture comprised of an encoder pathway (top left in Fig.~\ref{backbone_architecture}) and a supervised classifier module (bottom left in Fig.~\ref{backbone_architecture}), we propose a decoder module (top right in Fig.~\ref{backbone_architecture}) to be integrated with the architecture. The purpose of this module is to acts as an auto-encoder in order to generate attention maps in a latent space to preserve class-specific and domain-independent spatial information about the classification features throughout the network. 

Because pooling and convolution layers in the supervised classifier pathway reduce the spatial resolution of feature maps prior to the fully connected (FC) layers, important spatial relations will be lost in the forward path of the encoder module of the network. The loss of this class-specific and domain-independent spatial salient information is an important contributing factor in low performance of the traditional encoder-based architectures in the wild. 

The proposed decoder employs an unsupervised learning algorithm that is not presented with the ground truth or class labels of the training images. Salient information and spatial features are incorporated in the learning process by performing element-wise summation of the features from each stage of the encoder to the corresponding decoder stage. Eq.~\eqref{eq1} shows the proposed element-wise operation: 
\begin{equation}
   A = S_1\otimes C_n + U_n(\cdots S_{n-1}\otimes C_2 + U_2(S_n\otimes C_1 + U_1)))
   \label{eq1}
\end{equation}
where, $S_1,S_2,...S_n$ are the output of each stages of the encoder, while $U_1,U_2,..U_n$ symbolise the $2\times2$ upsampling layers of the decoder. $C_1,C_2,...,C_n$ are the convolution operations and maintain the same depth as the output of the corresponding upsampling layer. 

\begin{figure}[htp]
    \centering
    \includegraphics[width=\columnwidth]{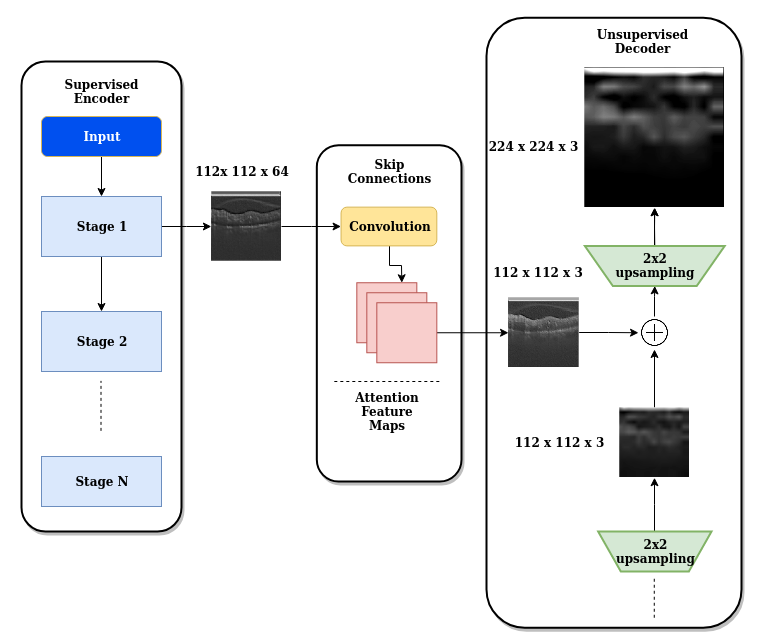}
    \caption{Attention Maps retrieved from each stages $S_1,S_2,....S_n$ added to the upsampled layers $U_1,U_2,....U_n$.}\vspace{-.2in}
    \label{attention_maps}
\end{figure}

Given any classification architecture, such as ResNet-50 or Xception, Eq.~\eqref{eq1} accounts for the variability of the number of stages to allow for appropriate element-wise operation. Fig.~\ref{attention_maps} shows the attention maps generated from these skip connections and the element-wise operation applied to the upsample layers. Spatial features important to different retina degenerations are amplified by fusing the spatial and depth information with output of the upsample layer. In order to retain this information, we back-propagate the loss for the unsupervised decoder and update gradients as a novel dual adaptive loss mechanism. 

\subsection{The Proposed Dual Adaptive Loss}
The architecture exploits and incorporates two different loss schemes to make the learning robust and efficient. Categorical cross-entropy is used for the supervised classifier given by $L_s$ equation in Eq.~\eqref{lossEq}. This loss function employs image labels $Y_i$. The second loss function employed in the unsupervised decoder pathway uses Mean-Squared Error (MSE) loss values given by the $L_u$ in Eq.~\eqref{lossEq}, to calculate the difference of the input and reconstructed output.  

\begin{equation}
    L_s = -\sum_{i=1}^{C} Y_i\log(Y'_i) \quad,\quad
    L_u = \frac{1}{N}\sum_{i}^{N}(P_i - P'_i)^2 
    \label{lossEq}
\end{equation}

In Equation.~\eqref{lossEq}, $Y_i$ signifies the true label or ground truth, $Y'_i$ symbolizes predicted output of the classifier, $C$ stands for the number of classes, $P_i$ is the input pixel value after normalization, $P'_i$ is the reconstructed output, and $N$ is the number of pixels in the image.  

For building the dual adaptive loss, we formulate an equation, given by Eq.~\eqref{totalLoss}, to adaptively interpolate between the supervised ($L_s$) and unsupervised ($L_u$) loss values. Here, $\phi$ is a user-specified hyper-parameter that can be tuned during training. The cumulative loss $L$ combines both $L_s$ and $L_u$ from equation \eqref{lossEq}, using a linear interpolation mechanism. The value of $\phi$ is between $0.0$ and $1.0$ . By default, and to evenly weight both the classifier and the unsupervised decoder output, we set $\phi$ to $0.5$. 

\vspace{-.1in}
\begin{equation}
    L = (\phi)*L_s + (1 -\phi)*L_u
    \label{totalLoss}
\end{equation}

By leveraging this weighted loss, we can prioritize which intermediate loss has more importance to update the gradient throughout the architecture. Larger $\phi$ values result in more importance be given to the $L_s$ loss related to the supervised classifier, while smaller $\phi$ values will result in more importance on $L_u$ loss related to the unsupervised decoder. Therefore, the hyper-parameter $\phi$ needs to be tuned based on how much the intermediate losses are decreasing. If the error starts to increase we change the $\phi$ value to reverse its effect. 
\vspace{-.1in}
\section{Experiments}
\subsection{Data-set and Preliminary Processing}
Bench-marking for the proposed architecture was done on two separate data-sets, \texttt{Srinivasan2014}~\cite{sri2014} and \texttt{OCT2017}~\cite{kermany2018identifying}. The \texttt{Srinivasan2014} dataset contains 3,231 images out of which 2,916 are use for training and 5-fold cross-validation and 315 are used for testing. The model with the best results for the test cases of the \texttt{Srinivasan2014} data-set is used for further testing on the unseen second data-set, i.e., \texttt{OCT2017}. The \texttt{OCT2017} consists of four distinct classes we take 250 cases of AMD and DME (in total 500 samples) for testing. We resize all samples for training and testing to $224\times224\times3$ resolution to adhere to the \texttt{Srinivasan2014} dataset format.

\begin{table}[hbp]
\caption{Test Results on Srinivasan2014\cite{sri2014} Dataset}
\begin{adjustbox}{width=\columnwidth,center}
\begin{tabular}{c cc cc cc }\toprule
\textbf{Architectures}& \multicolumn{2}{c}{\textbf{Accuracy}} & \multicolumn{2}{c}{\textbf{Specificity}} & \multicolumn{2}{c}{\textbf{Sensitivity}} \\ \toprule
ResNet50-v1 \cite{he2016deep}& \multicolumn{2}{c}{94.92}    & \multicolumn{2}{c}{97.46}    & \multicolumn{2}{c}{94.92}    \\ 
MobileNet-v2 \cite{sandler2018mobilenetv2}& \multicolumn{2}{c}{97.46}& \multicolumn{2}{c}{98.73}& \multicolumn{2}{c}{97.46}\\ 
OpticNet-71 \cite{kamran2019optic}& 
\multicolumn{2}{c}{100}      & \multicolumn{2}{c}{100}         & \multicolumn{2}{c}{100}         \\ \midrule
\textbf{\begin{tabular}[c]{@{}c@{}}Joint-Attention-Network\\ ResNet50-v1\end{tabular}}  & 
100 & $\uparrow$5.08  & 100 & $\uparrow$2.54 & 100 & $\uparrow$5.08 \\ 
\textbf{\begin{tabular}[c]{@{}c@{}}Joint-Attention-Network\\ MobileNet-v2\end{tabular}} & 
99.36 &   $\uparrow$7.46 & 99.68   & $\uparrow$0.95       & 99.36 &$\uparrow$1.9            \\ 
\textbf{\begin{tabular}[c]{@{}c@{}}Joint-Attention-Network\\ OpticNet-71\end{tabular}}  & 
99.68 &$\downarrow$0.32    & 99.84&$\downarrow$0.16& 99.68 &$\downarrow$0.32        \\ \bottomrule
\end{tabular}
\label{table1}
\end{adjustbox}
\end{table}

\begin{table}[tp]
\caption{Test Results on OCT2017\cite{kermany2018identifying} Dataset}
\label{tab:my-table}
\begin{adjustbox}{width=\columnwidth,center}
\begin{tabular}{c cc cc cc }\toprule
\textbf{Architectures}& \multicolumn{2}{c}{\textbf{Accuracy}} & \multicolumn{2}{c}{\textbf{Specificity}} & \multicolumn{2}{c}{\textbf{Sensitivity}} \\ \toprule
ResNet50-v1 \cite{he2016deep}& 83.40&$\downarrow$11.52    & 89.40 &  $\downarrow$8.06  & 83.40&$\downarrow$11.52    \\ 
MobileNet-v2 \cite{sandler2018mobilenetv2}& 93.80&$\downarrow$3.66& 96.70&$\downarrow$2.03 & 93.80& $\downarrow$3.66\\
OpticNet-71 \cite{kamran2019optic}& 
74.40 &$\downarrow$25.60 & 85.60 & $\downarrow$14.40 & 74.40 &$\downarrow$25.60         \\ \midrule
\textbf{\begin{tabular}[c]{@{}c@{}}Joint-Attention-Network\\ ResNet50-v1\end{tabular}}  & 
92.40 & $\uparrow$9.0  & 95.00 & $\uparrow$5.6 & 92.40 & $\uparrow$9.0 \\ 
\textbf{\begin{tabular}[c]{@{}c@{}}Joint-Attention-Network\\ MobileNet-v2\end{tabular}} & 
95.60 &   $\uparrow$1.8 & 97.1  & $\uparrow$0.4       & 95.60 &$\uparrow$1.8            \\ 
\textbf{\begin{tabular}[c]{@{}c@{}}Joint-Attention-Network\\ OpticNet-71\end{tabular}}  & 
77.40 &$\uparrow$3.0  & 89.00&$\uparrow$3.4& 77.40 &$\uparrow$3.0        \\ \bottomrule
\end{tabular}
\label{table2}
\end{adjustbox}
\end{table}

\vspace{-.25in}
\subsection{Performance Metrics}
For evaluation, we used three standard metrics on both the data-sets: Accuracy, Sensitivity  and Specificity. Both True Positive Rate (TPR) and True Negative Rate (TNR) are reported for \texttt{Srinivasan2014} and \texttt{OCT2017} data-sets. Performance metrics are, Accuracy $= \frac{1}{N}\sum TP$, Sensitivity $=\frac{1}{K}\sum\frac{TP}{TP + FN}$, and Specificity $=\frac{1}{K}\sum\frac{TN}{TN + FP}$.

\subsection{Training and Validation}
We worked with three different architectures to test and validate our results across two distinct data-set. First, we trained on the \texttt{Srinivasan2014} data-set with the original version of ResNet50-V1 \cite{he2016deep}, MobileNet-V2 \cite{sandler2018mobilenetv2}, and OpticNet-71 \cite{kamran2019optic}, using 5-fold cross validation. After choosing the best model we tested on the samples unseen by the architecture from a different dataset, i.e. \texttt{OCT2017}, consisting of 315 images. Next, we trained the Joint-Attention-Network of these three architectures on the \texttt{Srinivasan2014} samples and further tested on test-cases from both datasets. Table \ref{table1} reports all the results in terms of three metrics: Accuracy, Specificity and Sensitivity. Among the aforementioned methods in Table \ref{table1} only supervised classifier was used for training and testing. While for Joint-Attention-Network both the supervised classifier and unsupervised decoder were used. The unsupervised decoder used reconstructed output to compare with the input to calculate the loss hence not requiring any supervision or label annotation for training. 

In Table \ref{table2}, we extensively compare our three Joint-attention-network architectures with their original counterparts when tested on the \texttt{OCT2017} data-set. It should be noted that the \texttt{OCT2017} or any of its samples were not used for training or validation. We tested on 250 cases of AMD and 250 cases of DME images from this data-set. Joint-Attention-Network outperformed the original architectures and improved accuracy, precision, and recall in all cases. Henceforth, it is  evident that the proposed attention module is universally more robust across all scenario and setting. 

\subsection{Hyper-parameter Tuning}
For training Joint-Attention-Network we used Adam optimizer for both the supervised classifier and unsupervised decoder, with the initial learning rate set to  $\eta = 0.0001$. We used an adaptive learning scheme of updating the weights if the validation loss doesn't decrease for four consecutive epochs. The learning rate would change according to the equation $\eta = \eta * \kappa$, where $\kappa = 0.1$. Mini-batch size of $4$ and training was done for 30 epochs.

\section{Conclusions and Future Work}
In this paper, we propose a joint attention network that combines both supervised and unsupervised learning to make classification of retinal degeneration more robust and effective in the wild. Moreover, by incorporating the dual adaptive loss, our architecture addresses the issue of retaining spatial information throughout the network and updates the weights and the gradients accordingly. In future, we wish to extend our current research by improving the overall architecture and establishing a more robust weight updating mechanism. Finally, we hope that more salient features and oddities can be identified through this architecture, so that it can be used by clinicians for complex differential diagnosis.
\bibliographystyle{IEEEbib}
\bibliography{references}

\end{document}